\documentstyle[prl,aps]{revtex}
\begin{document}
\draft

\twocolumn
\narrowtext

{\large \bf Comment on ``Degenerate Wannier Theory for Multiple Ionization''
}

\vspace{3mm}

Recent Letter of Pattard and Rost \cite{PR} 
suggests new threshold laws for the processes of break up 
of a particle into several charged fragments, which, 
regretfully, is incorrect.
Previous studies, starting from the classical paper by 
Wannier \cite{Wannier} derived fragmentation cross section $\sigma$
in a power-law form 
\begin{eqnarray} \label{W}
\sigma(\epsilon \rightarrow 0) \sim \epsilon^\mu ~,
\end{eqnarray}
where $\epsilon$ is the energy excess above the break up threshold.
The primary task of the theory is to evaluate the threshold index $\mu$
for different multiparticle systems,
a comprehensive bibliography could be found in Ref.~\cite{KO}.
Pattard and Rost \cite{PR,PRerr} 
suggest the threshold law of novel functional form, 
namely, with an extra logarithmic $\epsilon$-dependence:
\begin{eqnarray} \label{flaw}
\sigma(\epsilon \rightarrow 0) \sim \epsilon^\mu \, 
|\ln \epsilon|^{-\nu} ~.
\end{eqnarray}

According to the originally idea of Wannier \cite{Wannier},
for the Coulomb interaction between the fragments
the threshold law is defined by the motion along
the {\it potential ridge}\/ as the constituent particles fly apart. 
This motion is unstable; survival on the ridge corresponds 
to the system fragmentation, whereas 
sliding from it means that the break up is not achieved.
The instability of the motion is characterized
by a set of Lyapunov exponents\/ $\lambda_i$.
The threshold index $\mu$ is expressed via 
{\it sum of all Lyapunov exponents} \cite{KO,PRerr}. 
In the present Comment we discuss 
the key result obtained by Pattard and Rost in \cite{PR},
and repeated in \cite{PRerr}, 
namely, the novel threshold law (\ref{flaw}). 

The standard approach to  derivation of threshold laws
relies on the analysis of classical trajectories in the vicinity of
the potential ridge where the equations of motion are
{\it linearized} over coordinates transversal to the ridge. 
The system is described by
a set of coupled linear equation that are differential over the time 
variable $t$. The time-dependence of the coefficients 
is eliminated by introducing the effective time 
$\tau$. The solutions of these equations $q(\tau)$ depend on 
the effective time as $\exp(\lambda_i \tau)$ where the set of 
$\lambda_i$ eigenvalues is found by solving the characteristic equation. 
The threshold index $\mu$ is expressed via $\lambda_i$, see for 
details Ref.~\cite{Wannier,KO} and bibliography therein.

Pattard and Rost \cite{PR,PRerr} essentially retain the traditional 
framework, but argue that the situation is changed  when
the Lyapunov exponents happen to be {\it degenerate}. 
They refer to the standard mathematical
textbook \cite{Sp} saying that ``If $n$ eigenvalues are degenerate,
the general solution contains additional terms 
$\tau^k \exp(\lambda \tau)$, $(k < n$)''. Further they relate 
the logarithm in the alleged coordinate time-dependence, 
\begin{eqnarray} \label{log}
q(\tau) = \tau^k \exp(\lambda \tau) \equiv 
\exp(\lambda \tau + k \ln \tau) 
\quad \quad (k<n) ~, 
\end{eqnarray}
to the logarithmic factor in the energy-dependence (\ref{flaw}). 

This reasoning is invalid.
Consider harmonic vibrations of a system with many degrees of freedom
(for example, a polyatomic molecule) around an equilibrium position.
Some eigenfrequencies
might be degenerate due to symmetry reasons, or accidentally.
The argumentation by Pattard and Rost fully applies to this
case; however, as universally known, the 
{\it logarithmic solutions}\/ (\ref{log}) in
fact do not emerge [see Eq.~(23.6) in Ref.~\cite{LL}
and subsequent discussion]. 
The reason is clear: {\it in the harmonic approximation 
all the normal modes are fully decoupled}\/ and 
hence a character of the time-dependence in each mode
does not depend on whether some other degenerate mode exists or not.

This argument is directly related to the motion 
in the vicinity of the potential ridge. The problem 
is described by the Hamiltonian that 
is quadratic both in transversal coordinates and momenta,
similar to the harmonic approximation for a polyatomic molecule.
The only difference is that the equilibrium is
unstable and some eigenfrequencies are complex-valued \cite{KO}.
Obviously this fact does not influence the functional form of
solutions of the same equations of motion which still reveal 
no {\it logarithmic terms}. 
The degenerate Lyapunov
exponents are related to the {\it different modes}\/ which
are {\it fully decoupled}, therefore the logarithmic solutions (\ref{log}),
being mathematically feasible for general linear differential 
equations, do not emerge in the physical applications concerned.

As a summary, the threshold law is given by Eq.~(\ref{W}) while the
modification (\ref{flaw}) is never valid.

This work has been supported by the Australian Research Council. 
V.~N.~O. acknowledges the hospitality of the staff of 
the School of Physics of UNSW where this work has been
carried out.

\vspace{2mm}

{M.~Yu.~Kuchiev and V.~N.~Ostrovsky 
}

{School of Physics, University of New South Wales,
Sydney 2052, Australia}

\vspace*{1.5mm}

\pacs{PACS numbers: 32.80.Fb, 34.80.Dp, 34.80.Kw, 31.15.Gy}

\end{document}